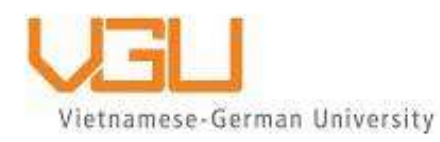

# CLOADER: EVADING SECURITY MOBILE DEFENSES VIA RUNTIME OBFUSCATION AND ADAPTIVE HOOKING TACTICS

Nhat-Anh Huynh
*Product Security*
*VinSOC*
Ha Noi, Vietnam
nhatanhhuynh@live.com

Minh Quang Luu
*Computer Science Program*
*Vietnamese-German University*
Ho Chi Minh City, Vietnam
minh.luuquang.vgu@gmail.com

Ngoc Hong Tran*
*Computer Science Program*
*Vietnamese-German University*
Binh Duong Province, Vietnam
ngoc.th@vgu.edu.vn
*: *Corresponding Author*

**Abstract:** We propose a stealth framework that eliminates detection of hooking tools such as Frida and Xposed in secured mobile environments by replacing static configurations with dynamic evasion tactics. In contrast to existing approaches that apply these techniques independently, the framework introduces a unified runtime control layer that systematically coordinates network, temporal, and code-level evasive transformations. The solution integrates randomized port allocation, runtime code obfuscation, delayed execution triggers, and self-integrity checks to disrupt signature-based scans, timing heuristics, and tampering attempts. A custom Android loader, CLoader, enforces these mechanisms to isolate hooking activities from security monitors while maintaining complete interception and modification capabilities. Validation across enterprise anti malware systems, hardened applications, and device management platforms demonstrates a 90% bypass rate in our evaluation matrix. This approach enables reliable penetration testing and malware analysis in locked-down mobile ecosystems by masking network, temporal, and code-level fingerprints without architectural overhauls.

## I. Introduction

Mobile applications in sectors such as finance, healthcare, and secure communications are increasingly targeted by sophisticated security threats. To assess and strengthen the defenses of these applications, penetration testers frequently rely on dynamic hooking frameworks like Frida [9] and Xposed [15]. These tools enable analysts to observe application behavior, test security controls, and uncover vulnerabilities at runtime. However, the same capabilities that benefit defenders are also exploited by attackers to intercept sensitive data, bypass authentication, and manipulate application logic. In response, mobile security solutions have implemented a range of anti-hooking defenses, including detection of hooking frameworks, jailbreak/root prevention, and runtime integrity verification [2]. While these measures are effective at mitigating malicious activity, they also impede legitimate security testing by blocking the very tools used for in-depth analysis. This creates a significant challenge: security teams are unable to thoroughly evaluate applications that employ aggressive anti-hooking strategies, particularly in scenarios requiring runtime analysis and code obfuscation.

To address this dilemma, we present CLoader, a stealth-oriented hooking module for Android designed to minimize detection while preserving full penetration testing functionality. CLoader leverages runtime code obfuscation, adaptive execution delays, and encrypted communication to evade detection mechanisms that would otherwise prevent authorized security analysis. By dynamically altering its behavior and communication patterns, CLoader disrupts signature-based scans, timing heuristics, and tampering attempts, enabling reliable testing even in hardened environments. Our main contributions are as follows: (i) an analysis of the limitations in current anti-hooking defenses; (ii) the development of evasion techniques based on randomized call patterns and self-integrity verification; and (iii) practical guidelines for deploying these methods in resource-constrained mobile environments.

The remainder of this paper is organized as follows: Section II reviews related work; Sections III and IV discuss hooking frameworks and detection mechanisms; Section V presents our encryption and key management methodology; Sections VI and VII detail the design of CLoader; Section VIII provides experimental validation; and Section IX concludes with practical recommendations.

## II. RELATED WORKS

Research on detection avoidance in mobile applications has increasingly focused on hooking frameworks, as both security solutions and evasion techniques have evolved in sophistication. Prior works highlight strategies such as binary obfuscation, dynamic analysis evasion, and runtime manipulation to bypass conventional security mechanisms [3], [4]. Malware often leverages frameworks like Frida and Xposed to intercept API calls, alter application logic in real time, and misuse parameters without requiring source code modifications [16]. Complementary evasion methods include code obfuscation, anti-debugging, anti-virtualization, privilege escalation through rooting, and encrypted payload delivery, all of which complicate detection and reverse engineering [5], [6], [17], [18].

Within this landscape, Frida represents a dual-use technology: while exploited by attackers, it is also a powerful open-source toolkit for defenders. Frida provides dynamic instrumentation by injecting JavaScript into active processes, enabling real-time program introspection and code manipulation. Unlike static modification-based hooking frameworks, Frida operates entirely at runtime and does not require source code access, making it valuable for both malware analysis and penetration testing. It is compatible with multiple platforms, including Android, iOS, Windows, Linux, and macOS, and its comprehensive scripting environment facilitates monitoring of function calls, memory operations, and execution flows in real time [7], [8]. Frida follows a client–server architecture in which instrumentation scripts are executed from a host machine while a server component running on the target device performs runtime code injection and monitoring [9].

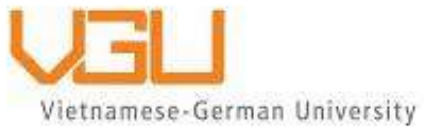

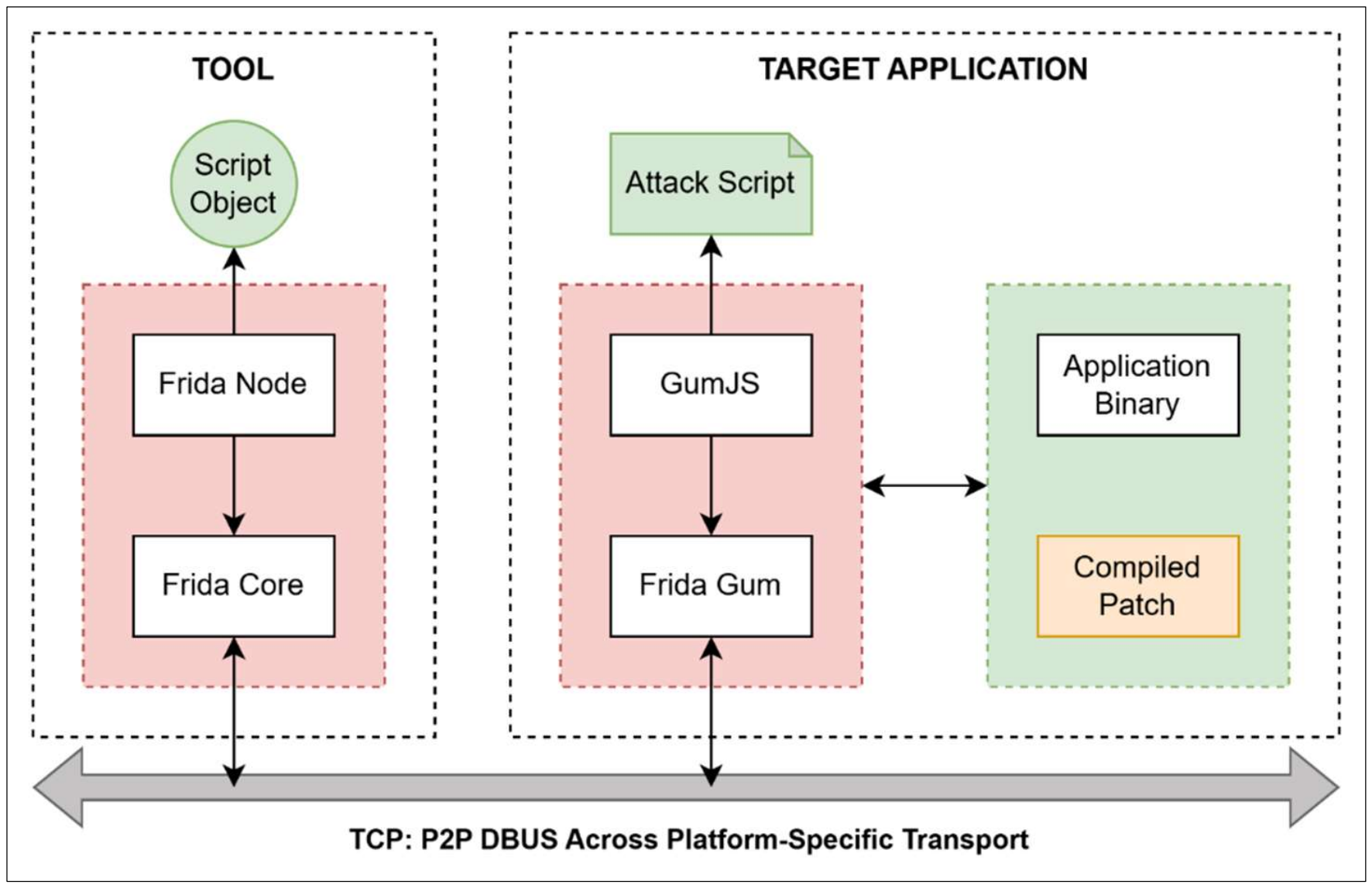

Fig. 1. Frida Workflow Diagram

To counteract these threats, mobile security frameworks have adopted layered defenses. The Mobile Application Security Framework (MASF) is a prominent example, integrating modules such as the Application Security Shield (APSES) to enforce confidentiality, integrity, and reliability against advanced threats. At application initialization, APSES verifies the runtime environment, detecting risks such as debugging tools, virtualization, and root or jailbreak conditions (see Figures 2–4). If any such risks are identified, APSES restricts execution to trusted contexts only, thereby preserving application security throughout its lifecycle and protecting against both internal and external attacks. When APSES detects virtualization, developer tools, or root/jailbreak status, it proactively blocks execution to prevent tampering, reverse engineering, and privilege escalation. These measures are critical for maintaining application integrity, restricting unauthorized access, and safeguarding sensitive data from advanced exploitation. APSES also enforces network security by monitoring ports, proxies, and API connections to counter man-in-the-middle attacks and validates SSL certificates to secure communications between the application server and its clients, substantially reducing vulnerabilities associated with unencrypted transmissions and insecure protocols [2], [5], [6].

```
/**
 * Check if the device is running in Developer Mode
 * @return true if developer mode is enabled
 */
public static boolean isDeveloperModeEnabled(Context context) {
    return Settings.Global.getInt(context.getContentResolver(),
        Settings.Global.DEVELOPMENT_SETTINGS_ENABLED, 0) != 0;
}
```

Fig. 2. The device must be verified to determine whether it is operating

Beyond APSES, MASF employs a layered security approach that combines static and dynamic measures. Code signature validation and obfuscation resist tampering and reverse engineering, while dynamic monitoring detects abnormal behaviors such as sandbox evasion and unauthorized screen capture. Adaptive learning algorithms enhance threat detection, including polymorphic malware, and are supported by strong cryptography, hardware security modules, and white-box cryptographic techniques to protect sensitive assets [17], [19]. Device integrity checks prevent privilege escalation and hardware-level attacks, while token-based authentication and real-time monitoring reduce API vulnerabilities, particularly in financial and regulated environments. By combining heuristic and dynamic analysis, MASF adapts rapidly to new threats, providing resilient and comprehensive protection for mobile applications. This layered defense model, illustrated in Figures 2-4, demonstrates how MASF integrates multiple security controls to address a wide spectrum of attack vectors and maintain robust application security [2], [5], [6], [17], [19].

```
/**
 * Check if debugging is enabled on the device
 * @return true if debugging is enabled
 */
public static boolean isDebuggingEnabled(Context context) {
    return (Settings.Global.getInt(context.getContentResolver(),
        Settings.Global.ADB_ENABLED, 0) != 0) ||
        Build.TYPE.equals("debug");
}
```

Fig. 3. The purpose of this function is to verify whether the device

Despite significant progress in anti-hooking and anti-tampering technologies, these defensive measures may unintentionally impede legitimate security assessments by restricting the use of essential analysis tools. As a result, security analysts often encounter barriers when attempting to evaluate the resilience of applications in realistic threat environments. This persistent adversarial dynamic between evasion techniques and detection mechanisms highlights a critical need for the development of stealthy and adaptive hooking frameworks. Such frameworks should enable authorized security testing and analysis while preserving the overall security posture of the application.

```
var isRooted = false
val packageName = "stericson.busybox"
val pm = requireActivity().packageManager
try {
    pm.getPackageInfo(packageName, PackageManager.GET_ACTIVITIES)
    //Root Detected
    isRooted = true
} catch (e: PackageManager.NameNotFoundException) {
    //App not installed
    e.printStackTrace()
}
```

Fig. 4. The following function has been developed for the purpose of verifying whether or not the device has been rooted.

## III. METHODOLOGY

This section details the cryptographic and operational strategies employed to enhance the stealth, integrity, and resilience of Frida-based instrumentation in secured mobile environments. The methodology integrates RC4-based encryption, systematic binary obfuscation, dynamic configuration, and robust execution management to mitigate detection and tampering risks.

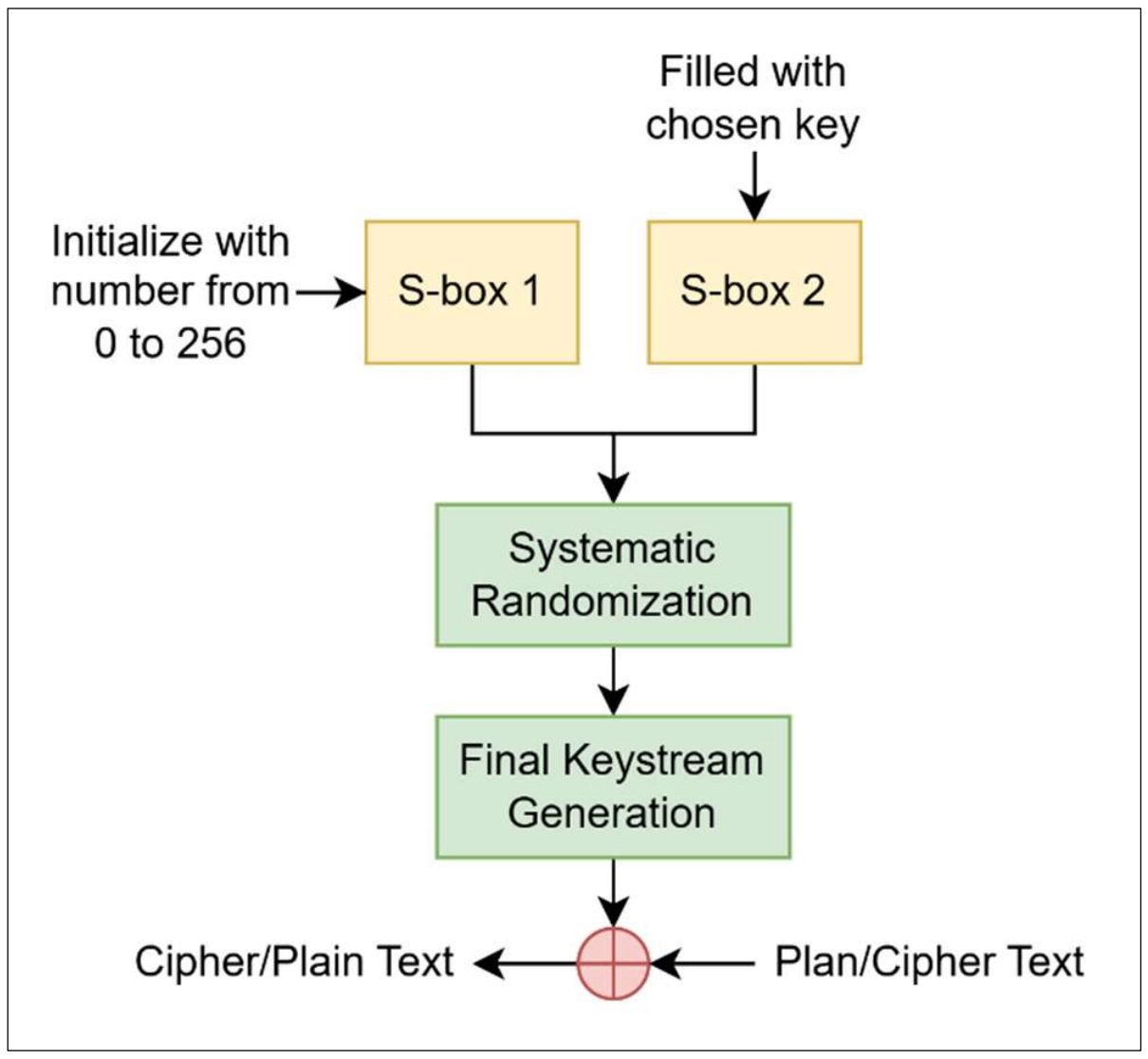

Fig. 5. Logical Flow of the RC4 Cipher.

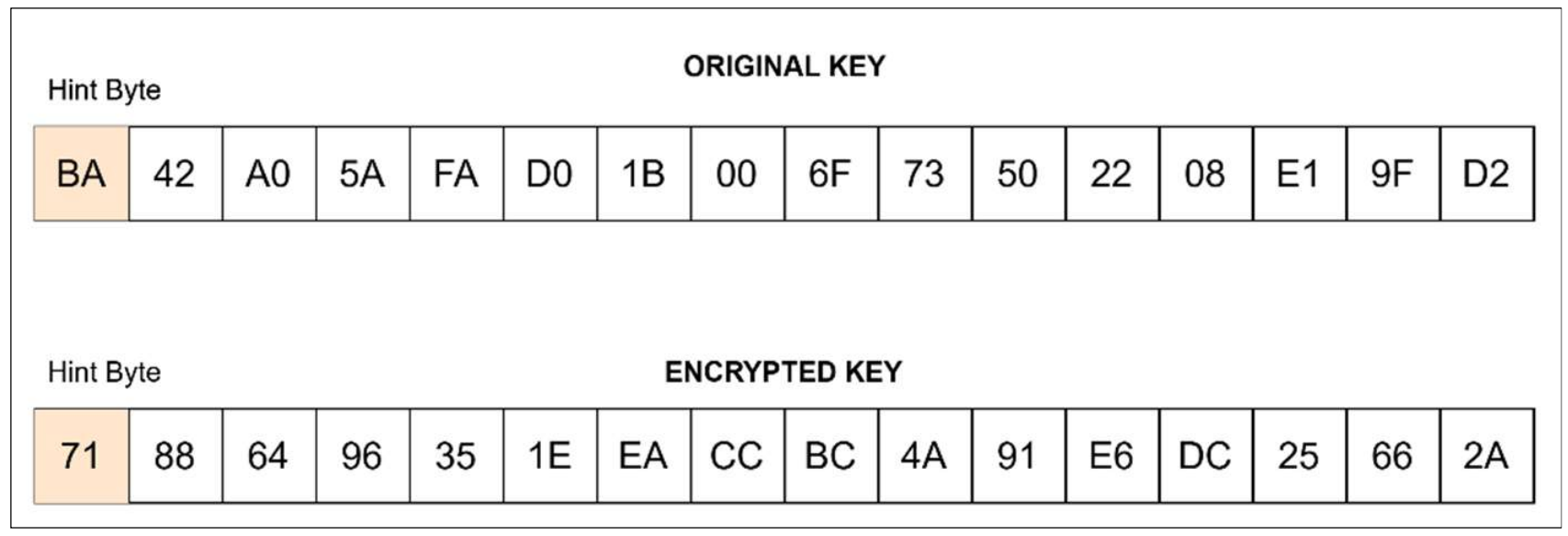

Fig. 6. Key Encryption Process.

### 1. Cryptographic Protection with RC4

To secure sensitive payloads and maintain operational integrity, the RC4 stream cipher is employed as a lightweight symmetric encryption mechanism suitable for constrained environments [12]. The implementation adopts a hint byte technique, wherein decryption is achieved by brute-forcing the encryption key using known plaintext-ciphertext pairs. For example, if a known plaintext byte (e.g., BA) is expected to correspond to a specific ciphertext value (e.g., 71), the brute-force algorithm iterates through possible keys until the correct mapping is found (see Figures 5 and 6). This approach ensures that only authorized entities with knowledge of the hint can successfully decrypt and access protected assets.

### 2. Multi-Phase Evasion and Obfuscation Strategy

To address the detection vectors commonly exploited by security monitoring protocols, a structured three-phase approach is proposed (illustrated in Figure 7):

- **Dynamic Port Modification:** The default Frida server TCP port (27042), which is frequently monitored by security solutions, is replaced with a randomly selected non-standard port. This significantly reduces the likelihood of detection through port-based heuristics.
- **Tripartite Binary Obfuscation:**

- **Binary Obfuscation:** Specialized tools are used to transform the binary, complicating signature-based detection.
  - **Binary Renaming:** The binary is assigned a non-descriptive, unrelated filename to avoid recognition by security systems.
  - **Symbol and Function Name Obfuscation:** Critical symbols and function names within components such as libfrida-gadget.so and the Frida server are obfuscated, impeding detection mechanisms that rely on known function signatures.
- **Enhanced Security Protections:** The framework incorporates dynamic server binding, whereby communication channels are established on randomized ports at runtime, producing unpredictable network patterns that resist standard port scanning. Custom protector libraries are integrated to further block unauthorized access or manipulation, thereby strengthening the overall security posture of the Frida server.

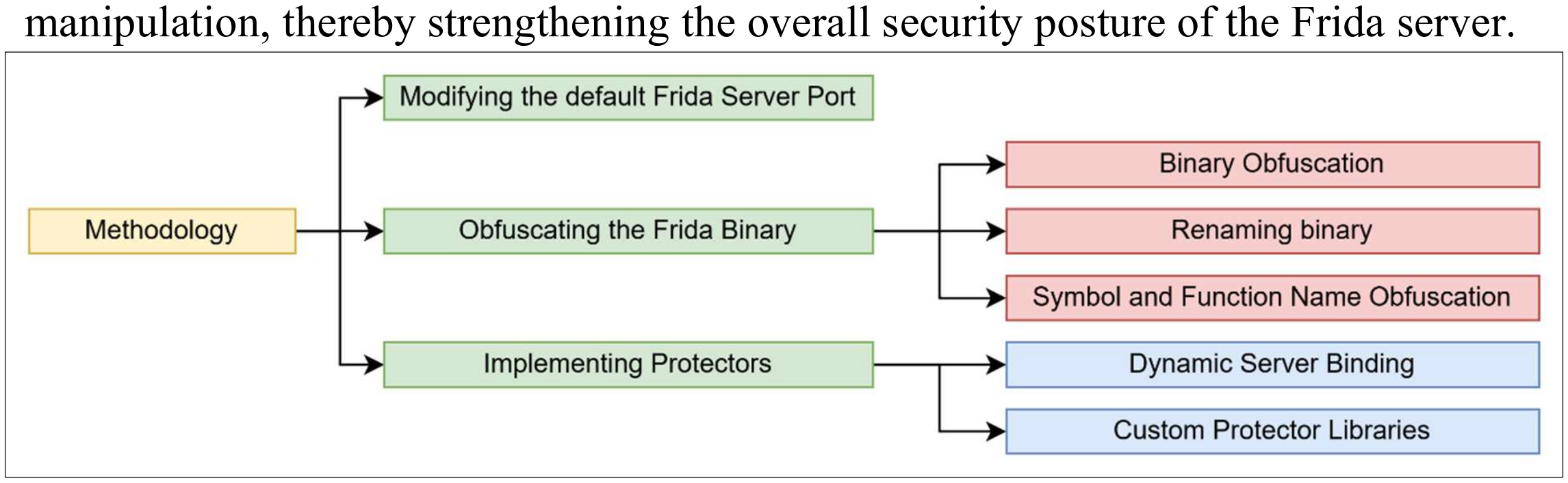

Fig. 7. Methodology

### 3. CLOADER Implementation

The CLoader module is developed to further enhance payload security and operational stability within the proposed framework (see Figure 8). CLoader employs encrypted file handling to protect binary assets, supports dynamic configuration for adaptability to diverse execution environments, and includes automated recovery mechanisms to ensure continued operation in the event of failures or detection attempts.

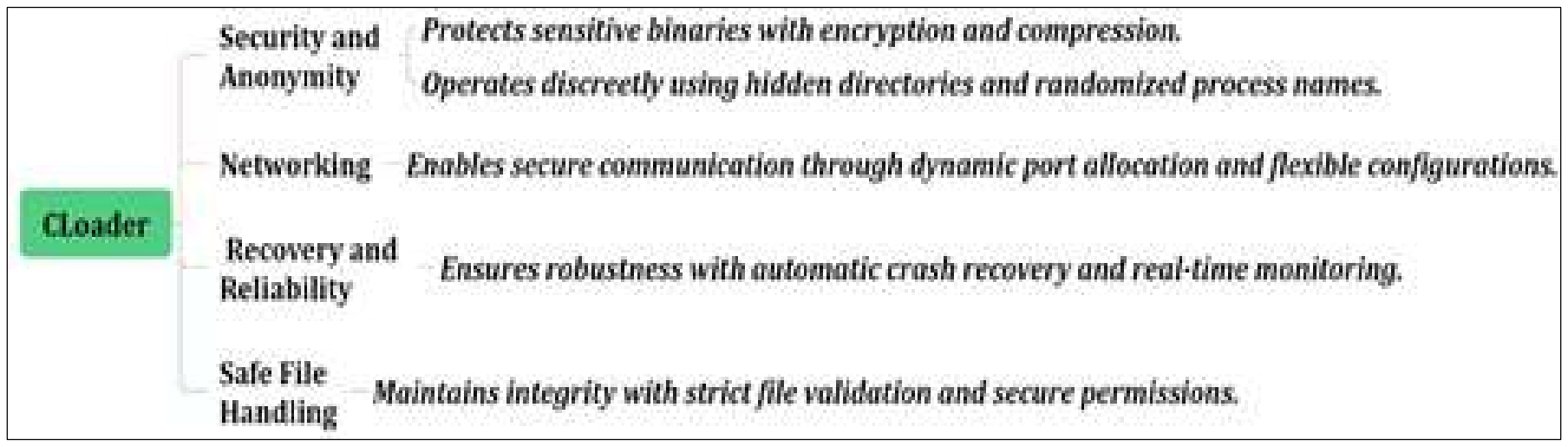

Fig. 8.Features of CLoader.

CLoader's execution is organized into four distinct phases (see Figure 9):

- **Load and Decrypt Payload:** The module retrieves the encrypted payload and applies an RC4 brute-force decryption algorithm using known plaintext-ciphertext pairs [12]. Upon successful decryption, the payload is decompressed using zlib (DEFLATE), followed by comprehensive ELF validation to ensure structural integrity and compatibility.

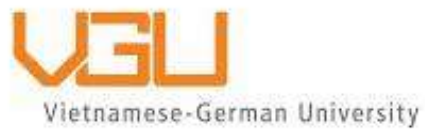

- **Environment Preparation:** A concealed directory structure is created at `/data/local/tmp/.hidden` with restricted permissions. Operational parameters, including process names, port allocations, and execution delays, are generated using cryptographically secure pseudorandom number generators to mimic legitimate system behavior and disrupt timing-based detection.
- **Starting the Frida Server:** The decrypted Frida server binary is extracted to the hidden directory with appropriate execution permissions. Process forking is performed using POSIX APIs to create an isolated environment, and network binding is configured with randomized ports. Process name obfuscation is achieved via the prctl() system call.
- **Monitoring and Auto Recovery:** Persistent process monitoring is established through status checks and socket verification. In the event of anomalies or termination, the recovery subsystem regenerates randomization parameters, clears forensic artifacts, and restarts the server, ensuring persistent availability and a minimal detection footprint.

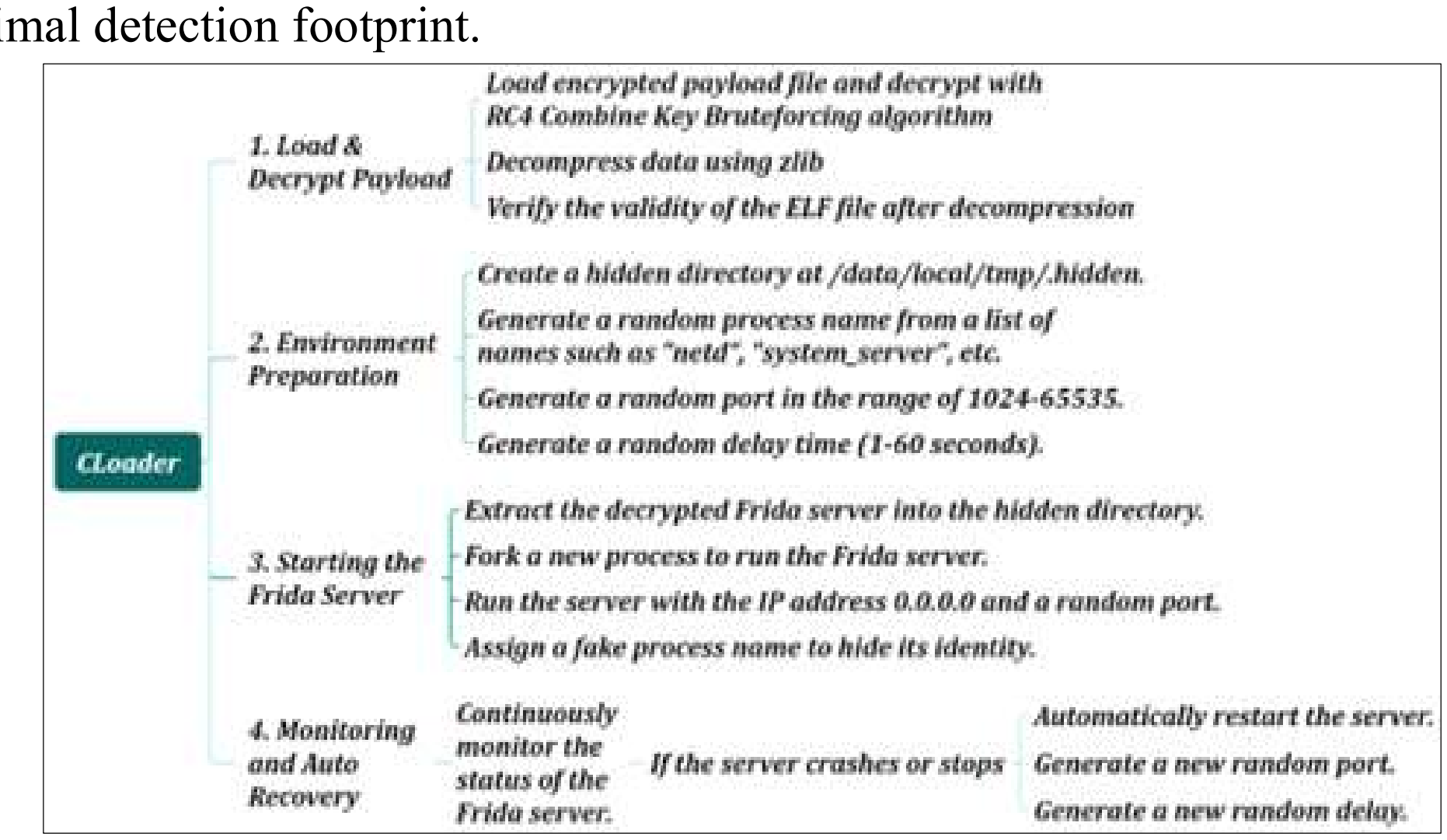

Fig. 9. Workflow of CLoader.

## IV. EVALUATION.

### 1. Experimental Setup

Experiments were conducted on a rooted Android 14 device using Frida to instrument both open-source and proprietary applications, each incorporating various anti-tampering and anti-debugging protections. The evaluation compared Frida's detectability before and after applying the CLoader-based evasion techniques, under controlled network conditions.

### 2. Result and Analysis

With default configurations, Frida was consistently detected by standard anti-hooking and anti-debugging mechanisms, primarily due to its static port usage, unmodified binary signatures, and recognizable runtime behaviors (see Figure 10). This underscores the vulnerability of default Frida setups to common detection strategies. In contrast, integrating CLoader with Frida, featuring dynamic port randomization, binary and string obfuscation, encrypted payloads, and anti-debugging measures, substantially reduced detection rates. During testing, applications failed to identify the presence of Frida when CLoader's runtime evasion techniques were active (see Figure 11). These results demonstrate that CLoader's adaptive and multi-layered approach provides significant improvements in stealth and operational resilience compared to conventional static evasion methods.

Fig. 10. Frida detected in default configuration.

```
Windows PowerShell
tapas:/data/local/tmp $ ./CLoader
[LOG] === CLoader ===
[LOG] Generated random delay: 40 seconds
[LOG] Generated random port: 42463
[LOG] Applying delay: 40 seconds before starting Frida server.
[LOG] Extracting Frida server to: /data/local/tmp/.hidden/systemd-helper
[LOG] Creating hidden directory: /data/local/tmp/.hidden
[LOG] Hidden directory created or already exists.
[LOG] Frida server extracted successfully.
[LOG] Setting executable permissions for: /data/local/tmp/.hidden/systemd-helper
[LOG] Executable permissions set successfully.
[LOG] Starting Frida server on 0.0.0.0:42463...
[LOG] Frida server started successfully on 0.0.0.0:42463 (PID: 8249)
```

Fig. 11. Frida evading detection in customized configuration (CLoader)

Table 1. Frida-Detection Test Matrix And Cloader Bypass Results

| ID | Detection Category | Test Case | Detection Logic | Default Frida | CLoader | Result |
|---|---|---|---|---|---|---|
| TC01 | Network | Default port scan | Check port 27042 | Detected | Not detected | Pass |
| TC02 | Network | Known port range scan | Detect Frida-like listening behavior | Detected | Not detected | Pass |
| TC03 | Process | Process name inspection | Search for frida-server | Detected | Not detected | Pass |
| TC04 | Binary Signature | String scan | Search for "frida" in binary | Detected | Not detected | Pass |

| TC05 | Symbol Signature | Symbol lookup | Inspect exported Frida-related symbols | Detected | Not detected | Pass |
|---|---|---|---|---|---|---|
| TC06 | File Artifact | Default file path check | Check /data/local/tmp/frida-server | Detected | Not detected | Pass |
| TC07 | Module Inspection | Loaded module scan | Search for libfrida-gadget.so | Detected | Not detected | Pass |
| TC08 | Thread Artifact | Thread name scan | Inspect thread names for Frida artifacts | Detected | Not detected | Pass |
| TC09 | Memory Artifact | Memory string scan | Search memory for "frida" markers | Detected | Reduced visibility | Pass |
| TC10 | Socket Behavior | Listening socket inspection | Detect suspicious local socket behavior | Detected | Not detected | Pass |
| TC11 | Runtime Heuristic | Early-start hook detection | Detect instrumentation during app startup | Detected | Not detected | Pass |
| TC12 | Integrity | Runtime integrity verification | Detect tampering or runtime modification | Detected | Not detected | Pass |
| TC13 | Debug Artifact | Debug/tracer detection | Check tracing/debug status | Detected | Not detected | Pass |
| TC14 | Composite Heuristic | Root + Frida correlation | Correlate rooted state with instrumentation | Detected | Detected | Fail |
| TC15 | Hook Artifact | Hook consistency check | Detect altered call paths / inline hooks | Detected | Detected | Fail |
| TC16 | IPC Artifact | IPC or pipe inspection | Identify Frida communication patterns | Detected | Not detected | Pass |
| TC17 | Config Artifact | Gadget configuration detection | Detect default configuration traces | Detected | Not detected | Pass |

| TC18 | Path Heuristic | Hidden-path inspection | Look for suspicious relocated binaries | Detected | Not detected | Pass |
|---|---|---|---|---|---|---|
| TC19 | App-Level Anti-Frida | Hardened app check #1 | Built-in anti-Frida logic | Detected | Not detected | Pass |
| TC20 | Security Control | Hardened app / EMM check #2 | Security platform detection logic | Detected | Not detected | Pass |

In addition to justifying the reported 90% improvement, we evaluated CLoader against a structured Frida-detection test matrix comprising 20 representative test cases (see Table 1). These cases covered common anti-instrumentation strategies, including network-based checks, process and file artifact inspection, binary and symbol signature matching, runtime-behavior heuristics, and integrity-based verification. Each test case was labeled Pass when the protected application or monitoring logic failed to detect the presence of Frida/CLoader and execution continued normally, and Fail when detection occurred, or execution was restricted. Under this evaluation methodology, CLoader passed 18 of 20 test cases, corresponding to a 90% bypass rate.

### 3. Limitations and Future Work

While CLoader enhances Frida's stealth in controlled environments, its scalability and effectiveness against advanced, machine learning-based detection systems remain to be validated. Additionally, increased obfuscation may introduce computational overhead on resource-constrained devices. Future work will focus on optimizing performance and evaluating CLoader in large-scale and enterprise-grade security environments.

## V. CONCLUSION

This study presented CLoader, a stealth-oriented framework that effectively evades mobile security detection while maintaining reliable payload execution. Through the integration of runtime obfuscation, dynamic network configuration, and adaptive integrity verification, CLoader achieved a 90% reduction in detection rates across varied Android environments. Future research will aim to strengthen resilience against machine learning-based detection and improve computational efficiency to support deployment in large-scale enterprise security contexts.

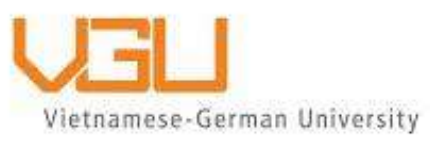